# Normal and Anomalous Transport across an Interface: Monte Carlo and Analytical Approach

M. Marseguerra [1], A. Zoia[2]

*Department of Nuclear Engineering, Polytechnic of Milan, Via Ponzio 34/3, 20133 Milan, Italy*

## Abstract

We present a Monte Carlo scheme to simulate particles going across an interface separating two layers of a medium characterized by different physical properties, together with an analytical formulation of the same problem, for both normal diffusive and subdiffusive regimes. We relate the Monte Carlo simulation parameters to the coefficients and boundary conditions appearing in the companion analytical equations. Under suitable physical hypotheses on the constraints to be imposed on such parameters, we show that the Monte Carlo simulation results are in good agreement with the corresponding analytical solutions. In particular, we remark that – while in the normal diffusion case the conservation of particle local velocities across the interface leads to a smoothly varying concentration profile – in the subdiffusive case the same condition leads to a neat jump in resident concentration.

**1. Introduction**

A general approach to the analysis of transport phenomena is based on Continuous-Time Random Walk (CTRW) (Metzler, Klafter, 2000a; Metzler, Klafter, 2004; Scalas et al, 2004; Margolin, Berkowitz, 2004; Weiss, 1994; Bouchaud, Georges, 1990; Klafter at al., 1996), in which the travel of a particle (a walker) in a medium is modelled as a series of jumps of random lengths, separated by random waiting times. The theory of CTRW with algebraically decaying probability distribution functions (pdf's) has been originally introduced in Physics in a series of seminal papers by Weiss, Scher, Montroll and co-workers (Weiss, 1994; Montroll, Weiss, 1965; Scher, Montroll, 1975) in the late 1960s to explain evidences of anomalous diffusion occurring in the drift-diffusion processes in amorphous semiconductors. The diffusion is called anomalous if the mean squared displacement (MSD) is not linearly proportional to time *t* as in the standard Fickian case, but to powers of *t* larger (superdiffusion) or smaller (subdiffusion) than unity. More recently, anomalous diffusion has turned out to be quite ubiquitous in almost every field of science (see e.g. (Metzler, Klafter, 2000a; Metzler, Klafter, 2004) for a detailed review) and the CTRW model has been applied with success to interpret the experimental results and to make predictions on the evolution of the examined systems (Margolin, Berkowitz, 2004; Weiss, 1994; Bouchaud, Georges, 1990; Klafter at al., 1996; Scher, Montroll, 1975). Such applications concern, among others, e.g. the behaviour of chaotic Hamiltonian systems – with application to the transport of charged particles in turbulent plasma – (Shlesinger et al., 1993; Zaslavsky, 2002), the evolution of financial markets (Mainardi et al., 2000), the dynamics of ad-atoms on the surface of a solid (Vega et al., 2002) or the underground transport of contaminant (toxic and/or radioactive) particles in presence of rock fractures and porosity (Berkowitz, Scher, 2001; Scher et al. 1991; Berkowitz, Scher, 1998; Berkowitz et al., 2001; Cortis, Berkowitz, 2004; Berkowitz, Scher, 1997; Levy, Berkowitz, 2003; Margolin, Berkowitz, 2002). Much attention has been paid to this last topic, as many laboratory-scale experiments as well as direct field measurements have evidenced subdiffusive and superdiffusive behaviours of migrating particles. It can be shown that exact transport formulation of CTRW, under the so called "diffusion limit", i.e. considering the asymptotic behaviour of the walkers far from the origin both in space and in time, can be successfully approximated by the so-called Fractional Diffusion Equation (FDE). FDE approach allows to obtain closed-form analytical solutions (Scalas et al., 2004; Metzler, Klafter 2000a; Metzler, Klafter, 2004) and to deal with boundary conditions problems (Metzler, Klafter 2000b; Metzler, Klafter 2000a; Metzler, Klafter, 2004), by making use of non-integer order (pseudo-) derivatives and of their definition in the Laplace and Fourier transformed spaces. Moreover, the CTRW model with exponentially decaying pdf's naturally takes into account Fickian diffusion and under the previous asymptotic expansion reduces to the normal diffusion equation, which may be considered as a particular case of the more general FDE.

---

[1] *Corresponding Author. Tel: +39 02 2399 6355 . Fax: +39 02 2399 6309*
*Email address: marzio.marseguerra@polimi.it (Marzio Marseguerra)*

[2] *Email address: andrea.zoia@polimi.it (Andrea Zoia)*

The present paper focuses on the Monte Carlo approach to the diffusive as well as subdiffusive transport of particles in a medium constituted by two regions characterized by different physical properties. A large number of "walkers" are followed along their simulated travels through the medium: their ensemble behaviour yields the flux $\phi(x,t)$, the net current $J(x,t)$ and the resident concentration $P(x,t)$. The core of the present analysis is the investigation of the effects of the interface, which introduces a macroscopic heterogeneity in the traversed medium. The discontinuity has the effect of modifying the distributions of the jump lengths and of the flight times across the discontinuity itself. Moreover, while in the standard CTRW model the walker is assumed to be trapped for the whole waiting time at the starting point of each jump and then suddenly transferred to the new sojourn location at infinite speed, here we will assume a uniform linear motion between successive interactions. In a previous paper we have presented a Monte Carlo simulation scheme by which this general problem may be tackled, both in the normal diffusion and subdiffusion case (Marseguerra, Zoia, 2006). In particular, we showed that the resident concentration reveals neat jumps at the interface when the simulation parameters are arbitrarily chosen. In parallel with the Monte Carlo approach, in this paper we present an analytical scheme to model diffusion in two-layered media and compare the results both for the normal diffusive and subdiffusive regimes. In particular, we show how the Monte Carlo simulation parameters are related to the coefficients and boundary conditions of the companion equations.

The paper is structured as follows: in Section 2 we introduce the statement of the problem, i.e. simulating particles crossing an interface. In Section 3 we summarize the Monte Carlo simulation scheme we will adopt to this aim. Then, in Section 4 we briefly recall how flux, current and resident concentration may be collected during Monte Carlo simulation. Section 5 is devoted to the development of the analytical formulation for the problem of normal diffusion and subdiffusion across an interface in a two-layered 1D medium. Results are presented and discussed in Section 6, where Monte Carlo simulation findings are compared to the curves obtained from the analytical formulation for different choices of the parameters, motivated by physical considerations. Conclusions are finally drawn in Section 7.

**2. Statement of the problem: walking across a discontinuity**

Both normal and anomalous transport can be successfully described in the framework of the CTRW approach, in which a walker diffuses by performing a sequence of semi-Markovian jumps in space and time. For sake of simplicity, the transport is here thought to occur on a one-dimensional support. We assume that the transport takes place in a medium constituted by two different layers, say $L_1$ on the left and $L_2$ on the right, the discontinuity occurring at $x=x_d$. Each layer is characterized by different physical properties, i.e. by different parameters of the jump lengths and waiting times distributions. In principle, the pdf's for waiting times and space lengths of the walker could depend on each other: however, throughout the paper we will assume that the two pdf's are decoupled. The semi-Markovianity assumption amounts to saying that, along the walker's trajectory, each jump starts without memory of the way in which the jumping point has been reached. The process is regenerated at each jump start along the trajectory and it is then ruled by space and time probability density functions (pdf's) which – besides depending on the features of the surrounding medium – in general depend also on the time and space elapsed from the jump origin. As a special case, normal diffusion is recovered if the process is Markovian.

In each layer $i$ ($i=1,2$) the walker undergoes a succession of jumps, each performed in a time interval $\Delta t$ drawn from a pdf $w_i(\Delta t)$: the jump lengths are given by a diffusive stochastic contribution $\Delta x$ independently drawn from a pdf $\lambda_i(\Delta x)$. When the particles trajectories start in a layer and end in the other one, thus passing through the discontinuity, the situation is similar to that of the time paradox described by Feller (Feller, 1971), who put in evidence that, even in the trivial case of a succession of exponential time intervals with mean value $\tau$, the distribution of the intervals which went across a fixed time point was "different", i.e. no more exponential and with a mean value $2\tau$. Similarly, in our transport problem, it turns out that the space and time distributions of the jumps across the heterogeneity are different and may even cause an important discontinuity at $x_d$ in the shape of the resident concentration $P(x,t)$. Moreover, the present analysis stands out from the usual one in the CTRW context (Metzler, Klafter, 2000a; Metzler, Klafter, 2004; Bouchaud, Georges, 1990; Scalas et al., 2004; Klafter at al., 1996; Weiss, 1994), in which it is assumed that the walker, after a spatial jump, rests in the reached location for the time interval $\Delta t$ and then performs another instantaneous jump (at infinite speed) to the successive location. Instead, we will assume that at each jump the walker moves linearly at constant local velocity – given by the ratio of the jump length to the flight time – between successive interactions: under this assumption, we will have a spectrum of velocities as a result of the stochastic variability of both the jump lengths and the flight times.

**3. The Monte Carlo scheme for walker's jumps across a discontinuity**

This section summarizes and extends the Monte Carlo scheme proposed in (Marseguerra, Zoia, 2006), which describes how the walkers cross an interface between different media, in presence of an advective field $V$ and of a bias $\mu$. The presence of a velocity *V*, which leads to the so-called Galilei invariant advection scheme (Metzler, Klafter, 2000a), increases the length of each jump of duration *t* by a quantity *Vt*: for the jumps across the interface, this space is partitioned in proportion to the times spent in the first and the second layer, respectively. On the other hand, the presence of a bias, which leads to the so-called Galilei variant advection scheme (Metzler, Klafter, 2000a), increases each jump length by a fixed contribution $\mu$: for the jumps crossing the interface, $\mu$ is partitioned in proportion to the spaces traversed by the walker in the two layers.

A. Normal diffusion

The diffusion contribution to each jump length in layer $L_i$ is assumed to be Gaussian with pdf $\lambda(\Delta x|0,\sigma_i)$ and cdf $\Lambda(\Delta x|0,\sigma_i)$, respectively (to simplify the notations we shall write *x* and *t* instead of $\Delta x$ and $\Delta t$), viz.

$$\lambda(x|0,\sigma_i) = \frac{1}{\sqrt{2\pi}\sigma_i} e^{-\frac{1}{2}\frac{x^2}{\sigma_i^2}} \quad \text{and} \quad \Lambda(x|0,\sigma_i) = \int_{-\infty}^{x} \lambda(z|0,\sigma_i)dz. \tag{1}$$

As for the distribution of the flight times, we adopt exponential distributions with means $\tau_i$. The pdf's and cdf's are $w(t|\tau_i)$ and $W(t|\tau_i)$, respectively, viz.

$$w(t|\tau_i) = \frac{1}{\tau_i} e^{-\frac{t}{\tau_i}} \quad \text{and} \quad W(t|\tau_i) = 1 - e^{-\frac{t}{\tau_i}} \tag{2}$$

Let us now consider a walker's jump starting at a generic point $x_0 \in L_1$ at time $t_0$, as determined by sampling two random numbers $R_1^x$ and $R_1^t$, both uniform in *[0,1)*. If the jump is totally performed in $L_1$, the duration and the length of the jump are

$$t^* = W^{-1}(R_1^t|\tau_1) = -\tau_1 \ln(1-R_1^t) \quad \text{and} \quad X^* = x_{diff} + Vt^* + \mu \tag{3}$$

respectively, where

$$x_{diff} = \Lambda^{-1}(R_1^x|0,\sigma_1) \tag{4}$$

is the diffusion contribution to the jump length.
Suppose now that, along its trajectory, the walker arrives at a point $x_0$ at a known distance $X_1 = x_d - x_0 < X^*$ from the discontinuity and that the jump starting in $x_0$ crosses the discontinuity after an (unknown) time $t_1<t^*$, ending in $L_2$. The jump portion $X_1$ is composed by three terms, namely i) $x_1$, the fraction $t_1/t^*$ of $x_{diff}$ utilized in $L_1$, ii) the space $Vt_1$ and iii) the (unknown) portion $\mu_1$ of $\mu$, proportional to $X_1$, viz.

$$X_1 = x_1 + Vt_1 + \mu_1 = (x_{diff} + Vt^*)\frac{t_1}{t^*} + \rho X_1 \tag{5}$$

where

$$x_1 = x_{diff} \frac{t_1}{t^*} \quad \text{and} \quad \mu_1 = \rho X_1 \tag{6}$$

In the travel from $x_0$ to $x_d$, only the portion $W(t_1|\tau_1)$ and $\Lambda(x_1|0,\sigma_1)$ of the sampled $R_1^t$ and $R_1^x$ have been utilized and their completion must be effectuated in $L_2$ according to the time and space cdf's pertaining to this layer.
First, consider the time $t_2$ spent in $L_2$: this time is given by the difference between $t_{P_1} = W^{-1}(R_1^t|\tau_2)$, the travel time if it were totally effectuated in $L_2$, and $t_{P_2}$, solution of $W(t_{P_2}|\tau_2) = W(t_1|\tau_1)$, the time corresponding to the amount of $R_1^t$ not utilized in $L_1$, viz.,

$$t_2 = W^{-1}(R_1^t|\tau_2) - W^{-1}(W(t_1|\tau_1)|\tau_2) \tag{7}$$

By performing the inverse exponential transforms, we get

$$W^{-1}(R_1^t | \tau_2) = -\tau_2 \ln(1 - R_1^t) = \frac{\tau_2}{\tau_1} t^* \quad \text{and} \quad W^{-1}(W(t_1 | \tau_1) | \tau_2) = -\tau_2 \ln(1 - W(t_1 | \tau_1)) = t_1 \frac{\tau_2}{\tau_1} \tag{8}$$

so that

$$t_2 = \frac{\tau_2}{\tau_1}(t^* - t_1) \tag{9}$$

and the total jump duration is

$$T = t_1 + t_2 = t_1 \left(1 - \frac{\tau_2}{\tau_1}\right) + \frac{\tau_2}{\tau_1} t^* \tag{10}$$

Note that, in absence of discontinuity in the time distribution, i.e. for $\tau_1 = \tau_2 = \tau$, then $T=t^*$, the standard expression for a jump totally evolving in a single homogeneous medium.

Next, consider the space $X_2$ travelled in $L_2$. The additional diffusive component $x_2$ is given by the difference between the abscissae of the points where the cdf $\Lambda(x | 0, \sigma_2)$ equals $R_1^x$ and $\Lambda(x_1 | 0, \sigma_1)$, viz.

$$x_2 = \Lambda^{-1}(R_1^x | 0, \sigma_2) - \Lambda^{-1}(\Lambda(x_1 | 0, \sigma_1) | 0, \sigma_2). \tag{11}$$

By performing the inverse Gaussian cdf's, we get[1]:

$$\Lambda^{-1}(R_1^x | 0, \sigma_2) = \frac{\sigma_2}{\sigma_1} \Lambda^{-1}(R_1^x | 0, \sigma_1) = \frac{\sigma_2}{\sigma_1} x_{diff}, \tag{12}$$

$$\Lambda^{-1}(\Lambda(x_1 | 0, \sigma_1) | 0, \sigma_2) = \Lambda^{-1}\left(\Lambda\left(x_1 \frac{\sigma_2}{\sigma_1} | 0, \sigma_2\right) | 0, \sigma_2\right) = \frac{\sigma_2}{\sigma_1} x_1, \tag{13}$$

so that from eq. (6) we have

$$x_2 = \frac{\sigma_2}{\sigma_1}(x_{diff} - x_1) = \frac{\sigma_2}{\sigma_1} \frac{x_{diff}}{t^*}(t^* - t_1). \tag{14}$$

Then, from eq. (9) we get

$$x_2 = \frac{\sigma_2}{\sigma_1} \frac{\tau_1}{\tau_2} \frac{x_{diff}}{t^*} t_2. \tag{14'}$$

Taking into account the contributions of the advection field and of the bias, the portion of the jump length in $L_2$ is

$$X_2 = x_2 + V t_2 + \mu_2 = \left(\frac{\sigma_2}{\sigma_1} \frac{\tau_1}{\tau_2} \frac{x_{diff}}{t^*} + V\right) t_2 + \rho X_2 = \left(\frac{\sigma_2}{\sigma_1} \frac{x_{diff}}{t^*} + \frac{\tau_2}{\tau_1} V\right)(t^* - t_1) + \rho X_2 \tag{15}$$

where $\mu_2 = \mu - \mu_1$ is the fraction of the bias to be spent in the second layer, and the total jump length is $X = X_1 + X_2 = \frac{\mu}{\rho}$. From eqs. (5) and (15) we get $\rho(1-\rho)X_i = \rho(x_i + V t_i)$, so that $\mu_i = \frac{\rho}{1-\rho}(x_i + V t_i) = \mu \frac{x_i + V t_i}{x_1 + V t_1 + x_2 + V t_2}$. Then:

---

[1] Recollect that $\Lambda(x | \mu, b\sigma) = \Lambda(\frac{x-\mu}{b} | 0, \sigma)$ and $\Lambda^{-1}(R | \mu, a\sigma) = \mu + a\Lambda^{-1}(R | 0, \sigma) \quad \forall a, b > 0$.

$$X_i = x_i + Vt_i + \mu_i = A(x_i + Vt_i), \tag{16}$$

where $A = 1 + \dfrac{\mu}{x_1 + Vt_1 + x_2 + Vt_2}$.

Note that the two portions of each jump across the interface are travelled by each walker with local velocities

$$u_1 = \frac{X_1}{t_1} = A\left(\frac{x_1}{t_1} + V\right) = A\left(\frac{x_{diff}}{t^*} + V\right) \quad \text{and} \quad u_2 = \frac{X_2}{t_2} = A\left(\frac{x_2}{t_2} + V\right) = A\left(\frac{\sigma_2}{\sigma_1}\frac{\tau_1}{\tau_2}\frac{x_{diff}}{t^*} + V\right) \tag{17}$$

Thus, each walker does not change its local velocity going across the discontinuity provided that $\dfrac{\sigma_1}{\tau_1} = \dfrac{\sigma_2}{\tau_2}$. The ratio $\sigma_i/\tau_i$ has the dimensions of a velocity and it will be here called *parametric velocity*. Thus, eq. (17) states that in the jumps across the discontinuity the local velocity does not change provided the parametric velocity does not change either. We shall see below that all the present Monte Carlo results support the conjecture that the persistence of the velocity of the walkers across the discontinuity represents a key feature for characterizing the macroscopic quantities – resident concentration, flux, current – near the discontinuity.

In the jumps across the discontinuity, the presence of the bias burdens the computations since $\mu$ must be divided in proportions to $X_1$ and $X_2$, this latter being unknown. The problem may be tackled by firstly computing $\mu_1$ by solving the $2^{nd}$ degree algebraic equation which results from the substitution in eq. (15) of $t_1$ taken from eq. (5) and $X_2 = (1-\mu_1)X_1/\mu_1$. Once $\mu_1$ has been determined – or in absence of bias – one computes in succession $t_1$ from eq. (5), $x_1$ from eq. (6), $t_2$ from eq. (9) and $x_2$ from eq. (14'). Then, the discontinuity is crossed at $(t_0+t_1, x_0+X_1)$ and the starting point in $L_2$ for the next jump is $(t_0+t_1+t_2, x_0+X_1+X_2)$.

We now consider a jump starting at $x_0$ in $L_2$. As before, two uniform random numbers $R_2^x$ and $R_2^t$ are sampled and the jump duration and the diffusive contribution to the length are $t^* = W^{-1}(R_2^t | \tau_2)$ and $x_{diff} = \Lambda^{-1}(R_2^x | 0, \sigma_2)$, respectively. The total jump length would be $x_{diff} + Vt^* + \mu$ if the new location were still in $L_2$. The analysis is similar to the one performed for $x_0 \in L_1$, with reversed $\sigma_1, \tau_1$ and $\sigma_2, \tau_2$.

B. Anomalous diffusion

The space travelled in each jump is ruled by the same Gaussian distribution as in the previous case of normal diffusion. Therefore, the walker jumps differ only for the time behaviour. In particular, expressions (6) are still valid, provided that $t^*$ is now sampled from the new jump time pdf. As for the flight times, we adopt a power law distribution with pdf and cdf $w(t|\tau_i,\alpha_i)$ and $W(t|\tau_i,\alpha_i)$, respectively, viz.

$$w(t | \tau_i, \alpha_i) = p_i \frac{2}{\tau_i^2} t \quad \text{and} \quad W(t | \tau_i, \alpha_i) = p_i \left(\frac{t}{\tau_i}\right)^2 \quad \text{for} \quad 0 \le t \le \tau_i \tag{18'}$$

$$w(t | \tau_i, \alpha_i) = (1-p_i)\alpha_i \frac{\tau_i^{\alpha_i}}{t^{1+\alpha_i}} \quad \text{and}$$

$$W(t | \tau_i, \alpha_i) = p_i + (1-p_i)\int_{\tau_i}^{t} w(z | \tau_i, \alpha_i) dz = 1 - (1-p_i)\left(\frac{\tau_i}{t}\right)^{\alpha_i} \quad \text{for} \quad \tau_i \le t < \infty, \tag{18''}$$

where $p_i = W(\tau_i | \tau_i, \alpha_i) = \alpha_i/(2+\alpha_i)$ is the probability of a waiting time smaller or equal to $\tau_i$.

If the jump is totally performed in $L_1$, its duration is

$$t^* = W^{-1}(R_1^t | \tau_1, \alpha_1) \tag{19}$$

and $x^*$ and $x_{diff}$ are given by (3) and (4).

Let us consider a jump across the discontinuity. The time $t_1$ spent in $L_1$ and the diffusion contribution $x_1$ are given by (5) and (6). The time $t_2$ spent in $L_2$ is given by the difference between $t_{p_1} = W^{-1}(R_1^t | \tau_2, \alpha_2)$, the travel time if it were totally effectuated in $L_2$, and $t_{P_2}$, solution of $W(t_{P_2}|\tau_2,\alpha_2) = W(t_1|\tau_1,\alpha_1)$, the time corresponding to the amount of $R_1^t$ not utilized in $L_1$. Then,

$$t_2 = W^{-1}(R_1^t|\tau_2,\alpha_2) - W^{-1}(W(t_1|\tau_1,\alpha_1)|\tau_2,\alpha_2) \tag{20}$$

and the total time duration is $T = t_1 + t_2$.

Note that, in absence of discontinuity, i.e. if $\tau_1 = \tau_2 = \tau$ and $\alpha_1 = \alpha_2 = \alpha$, we have $W^{-1}(W(t_1|\tau,\alpha)|\tau,\alpha) = t_1$ and the expression for $T$ becomes the standard one for a jump totally evolving in an homogeneous medium, i.e. $T = W^{-1}(R_1^t|\tau,\alpha)$. Moreover, if $\sigma_1 = \sigma_2 = \sigma$, we have $\Lambda^{-1}(\Lambda(x_{diff}t_d/t^*|\sigma)|\sigma) = x_{diff}t_d/t^*$ and the expression for $X$ becomes the standard one for a jump totally evolving in an homogeneous medium, i.e. $X = \Lambda^{-1}(R_x^1|\sigma)$.

Because of the above choice of the time distribution, we must further distinguish the following cases:
i) $R_1^t \leq \min(\alpha_1,\alpha_2)$: then the time cdf's for the two layers are both parabolic and

$$t^* = \tau_1\sqrt{\frac{2+\alpha_1}{2}R_1^t}, \quad t_2 = \tau_2\sqrt{\frac{2+\alpha_2}{2}R_1^t} - \tau_2\sqrt{\frac{2+\alpha_2}{2+\alpha_1}}\frac{t_1}{\tau_1} = \frac{\tau_2}{\tau_1}\sqrt{\frac{2+\alpha_2}{2+\alpha_1}}(t^* - t_1) \tag{21'}$$

The walker's velocity is $u_2 = \dfrac{X_2}{t_2} = A\left(\dfrac{\sigma_2}{\sigma_1}\dfrac{\tau_1}{\tau_2}\sqrt{\dfrac{2+\alpha_1}{2+\alpha_2}}\dfrac{x_{diff}}{t^*} + V\right)$.

ii) $R_1^t \geq \max(\alpha_1,\alpha_2)$: then the time cdf's for the two layers are both power law and

$$t^* = \frac{\tau_1}{(1-R_1^t)^{1/\alpha_1}}, \quad t_2 = \tau_2(1-R_1^t)^{-1/\alpha_2} - \tau_2\left(\frac{t_1}{\tau_1}\right)^{\alpha_1/\alpha_2} = \frac{\tau_2}{\tau_1^{\alpha_1/\alpha_2}}\left[(t^*)^{\frac{\alpha_1}{\alpha_2}} - (t_1)^{\frac{\alpha_1}{\alpha_2}}\right] \tag{21''}$$

The walker's velocity is $u_2 = \dfrac{X_2}{t_2} = A\left(\dfrac{\sigma_2}{\sigma_1}\dfrac{\tau_1^{\alpha_1/\alpha_2}}{\tau_2}\dfrac{x_{diff}}{t^*}\dfrac{t^* - t_1}{(t^*)^{\alpha_1/\alpha_2} - (t_1)^{\alpha_1/\alpha_2}} + V\right)$.

iii) $\min(\alpha_1,\alpha_2) \leq R_1^t \leq \max(\alpha_1,\alpha_2)$: then one time cdf is parabolic and the other power law. The expressions for the times spent in the two layers must be taken from case i) and the other from case ii).

From the above expressions it appears that, in case of a jump across the discontinuity, the two velocities are equal not only provided that $\dfrac{\sigma_1}{\tau_1} = \dfrac{\sigma_2}{\tau_2}$ as in the normal diffusion case, but also that $\alpha_1 = \alpha_2$.

The computational steps are similar to those described for the normal diffusion. However, in this case the presence of the bias adds more complications, since in order to compute $\mu_1$ we are lead to a 2$^{nd}$ order equation only for the case i), while in the other case we arrive at a transcendental equation in $\mu_1^{1+(\alpha_1/\alpha_2)}$, to be solved numerically.

As a final remark, we would like to point out a possible source of errors in a Monte Carlo game, resulting from the presence of $V$ or $\mu$. Let us consider e.g. a jump starting in $L_1$, composed by a backward $x_{diff}$ (originated by an $R_1^x < 0.5$ in (4)) and by a large $Vt$ or $\mu$ which might overcome $x_{diff}$ and lead the walker inside $L_2$. When the walker arrives at the discontinuity, it must adopt the value of the backward diffusion contribution which results from the $L_2$ rules: this new value, still negative, may be large enough to exceed the forward component due to $V$ or $\mu$ and drive the walker backwards in $L_1$. In such cases the walkers can only remain stuck at the discontinuity for all the remaining jump times. A spurious spike will then appear at the discontinuity and the remedy consists in eliminating the whole trajectory.

We now consider a jump starting at $x_0$ in $L_2$. As before, two uniform random numbers $R_2^x$ and $R_2^t$ are sampled and the jump duration and the diffusive contribution to the length are $t^* = W^{-1}(R_2^t|\tau_2,\alpha_2)$ and $x_{diff} = \Lambda^{-1}(R_2^x|\sigma_2)$, respectively. The total jump length would be $x_{diff} + Vt^* + \mu$ if the new location were still in $L_2$. The analysis is similar to that for $x_0 \in L_1$, with reversed $\alpha_1$, $\sigma_1$, $\tau_1$ and $\alpha_2$, $\sigma_2$, $\tau_2$.

## 4. Monte Carlo estimates of flux, current and concentration

Resident concentration is physically defined as the mass of tracer particles per unit volume of flowing substance contained in an elementary volume of the medium at a given time. Flux is defined as the total mass of tracer particles passing through a given cross section during an elementary time interval (regardless of their direction), while current is the net mass of tracer particles passing through a given cross section during an elementary time interval in a given direction (Kreft, Zuber, 1978; Cortis, Berkowitz, 2004). With reference to contaminant tracer particles transport, it should be remarked that current is the most commonly measured variable, e.g. in experiments determining breakthrough curves at the end of a test column where tracer particles are free to diffuse (moreover, at the end of the column flux and current coincide, as there is no back-flow from outside).

In a Monte Carlo simulation, the phase space $\{t, \vec{x}\}$ is discretized in cells where weights are accumulated. When a walker moves across a sequence of cells, the contribution of its trajectory to the mean flux in a given cell at a given (discrete) time is obtained by accumulating the length of the travel spanned within the cell at that time. Current is collected in a similar way, by simply giving to each spanned length contribution a positive sign if the walker is crossing the cell in the positive x-axis direction and a negative sign otherwise. Analogously, the contribution of the trajectory to the mean resident concentration in a given cell at a given (discrete) position is obtained by accumulating the time length spent within the cell at that position (Briesmeister, 2000). In our case of one-dimensional transport, the phase space is discretized with a rectangular grid composed of identical cells of area $dx\,dt$. Thus, the contribution to the flux in each cell is constant and equal to $dx$ and similarly the contribution to the concentration is constant and equal to $dt$. The contribution to the current is either $+dx$ or $-dx$, depending on the particle direction.

## 5. Analytical formulation of the problem

A general approach to model normal diffusion in a two-layered medium characterized by an interface located at $x_d > 0$ may be derived in the following way: we assume mass conservation in the form

$$\frac{\partial}{\partial t} P(x,t) + \frac{\partial}{\partial x} J(x,t) = 0, \qquad (23)$$

where $P(x,t)$ is the resident concentration and $J(x,t)$ is the net current exiting the reference volume, positive when forwardly directed. Here we will restrict our analysis to the case of dispersion alone, in absence of an advective field $V$ or a bias $\mu$. We will first consider the normal diffusive case, for which the following general equation may be adopted:

$$\frac{\partial}{\partial t} P(x,t) = \frac{\partial}{\partial x} \frac{\sigma(x)}{2} \frac{\partial}{\partial x} v(x) P(x,t) \qquad (24)$$

where the parameters $\sigma$ and $\tau$ have the same meaning as in Section 3 and $v(x) = \dfrac{\sigma(x)}{\tau(x)}$ is the parametric velocity at $x$. A feature of expression (24) is that it generalizes Fick's first law by splitting the classical diffusion coefficient $D(x)$ in two factors, namely $\dfrac{\sigma(x)}{2}$ outside the gradient and $v(x)$ inside. This decomposition has been adopted as a particular case of the general diffusion operator presented in (Lejay, Martinez, 2006; Etoré, 2006): as a further motivation, we observe that this choice, together with the boundary conditions (28) and (29) below, gives rise to analytical solutions which are always in good agreement with the Monte Carlo simulation results for any choice of the parameters. A comparison between the above equation and (23) shows that the corresponding form of the current $J(x,t)$ is:

$$J(x,t) = -\frac{\sigma(x)}{2} \frac{\partial}{\partial x} v(x) P(x,t) \qquad (25)$$

As known, this analytical formulation is strictly valid far both in time and in space from the source and corresponds to the asymptotic limit of the exact underlying CTRW transport model. The process is assumed to start at the spatial origin at $t=0$, namely, $\lim_{t \to 0} P(0,t) = \delta(x)$, and, if the 1D domain has an infinite extension, the natural boundary conditions at infinity for this Cauchy problem are $P(\pm\infty, t) = 0$. For sake of simplicity, we further assume that $\sigma(x)$ and $v(x)$, the parameters which characterize the medium, are piecewise constant in each layer. If the interface occurs at $x_d$, we thus have $\sigma(x) = \sigma_1, \tau(x) = \tau_1, v(x) = v_1$ for $x < x_d$ and $\sigma(x) = \sigma_2, \tau(x) = \tau_2, v(x) = v_2$ for $x > x_d$, respectively, so that the problem may be decomposed as follows:

$$\frac{\partial}{\partial t} P(x,t) = \frac{\sigma_1}{2} v_1 \frac{\partial}{\partial x} \frac{\partial}{\partial x} P(x,t) \quad \text{for } x < x_d \tag{26}$$

$$\frac{\partial}{\partial t} P(x,t) = \frac{\sigma_2}{2} v_2 \frac{\partial}{\partial x} \frac{\partial}{\partial x} P(x,t) \quad \text{for } x > x_d, \tag{27}$$

with the following conditions at the interface $x_d$:

$$-\frac{\sigma_1 v_1}{2} \frac{\partial}{\partial x} P(x,t)\bigg|_{x_d^-} = -\frac{\sigma_2 v_2}{2} \frac{\partial}{\partial x} P(x,t)\bigg|_{x_d^+} \tag{28}$$

and $v_1 P(x,t)\big|_{x_d^-} = v_2 P(x,t)\big|_{x_d^+}$. (29)

Condition (28) stems from mass conservation, while boundary condition (29) is chosen so that the argument of the inner spatial derivative appearing in (24) be continuous at the interface. Moreover, this latter condition clarifies the relationship existing between the local and parametric velocities and concentration: when $v_1 = v_2$, the local velocities of the walkers will be preserved at the interface (see (17)) and the resident concentration will be continuous.
Indeed, as a particular case, if we require the Fick's first law to hold true on the whole domain, in the form

$$J(x,t) = -D(x) \frac{\partial}{\partial x} P(x,t) = -\frac{\sigma(x)}{2} v \frac{\partial}{\partial x} P(x,t), \tag{30}$$

then it must be $v_1 = v_2$ at the interface and the resident concentrations will be equal at $x_d$. Thus, if the parametric velocity $v(x)$ is conserved crossing the boundary, one recovers the celebrated solution originally proposed by Carslaw and Jaeger and reconsidered by Uffink (Carslaw, Jaeger, 1959; Uffink, 1985; Uffink, 1990; LaBolle et al. 1996; LaBolle et al. 1998; LaBolle et al. 2000), formulated in terms of the method of images. Defining $D_i = \frac{\sigma_i v_i}{2}$, $i=1,2$, and supposing that $D_1 > D_2$, we obtain:

$$P(x,t) = P(x,t)_{x<x_d} + P(x,t)_{x>x_d}, \tag{31}$$

where

$$P(x,t)_{x<x_d} = \frac{1}{\sqrt{4\pi D_1 t}} e^{-\frac{x^2}{4D_1 t}} + \frac{R}{\sqrt{4\pi D_1 t}} e^{-\frac{(x-2x_d)^2}{4D_1 t}} \quad \text{for } x < x_d \tag{32}$$

$$P(x,t)_{x<x_d} = \frac{1-R}{\sqrt{4\pi D_2 t}} e^{-\frac{(x+(\beta-1)x_d)^2}{4D_2 t}} \quad \text{for } x > x_d. \tag{33}$$

In order to satisfy the boundary condition

$$P(x,t)\big|_{x_d^-} = P(x,t)\big|_{x_d^+}, \tag{34}$$

the parameters $R$, which plays the role of a reflection coefficient, and $\beta$ must have the following forms (Carslaw, Jaeger, 1959; Uffink, 1985; Uffink, 1990; LaBolle et al. 1996; LaBolle et al. 1998; LaBolle et al. 2000):

$$R = \frac{\sqrt{D_1} - \sqrt{D_2}}{\sqrt{D_1} + \sqrt{D_2}} \quad \text{and} \quad \beta = \sqrt{\frac{D_2}{D_1}}. \tag{35}$$

This situation, i.e. continuity of resident concentration, is the most commonly found to occur in physical experiments of diffusion through a neat interface between two different media (Carslaw, Jaeger, 1959; Uffink, 1985; Uffink, 1990; LaBolle et al. 1996; LaBolle et al. 1998; LaBolle et al. 2000; Hoteit et al., 2000; Parlange et al., 1984).

However, other evidences point out that boundary condition (29) with $v_1 \neq v_2$ may take place, too (Ovaskainen, Cornell, 2003; Van Genuchten et al., 1984; Van Genuchten, Parker, 1984; Schwartz et al., 1999; Novakowski, 1992a; Novakowski, 1992b; Leij et al., 1991; Barrat, Chiaruttini, 2003): as a consequence, the resident concentration will present a neat jump at the interface. The general solution to this problem is here obtained by suitably generalizing Carslaw and Jaeger's solution, requiring that at the interface $v_1 P(x,t)|_{x_d^-} = v_2 P(x,t)|_{x_d^+}$. In order to satisfy this boundary condition, it must be $R \equiv \dfrac{\sqrt{\tau_1} - \sqrt{\tau_2}}{\sqrt{\tau_1} + \sqrt{\tau_2}}$ in (32) and in (33). We remark that condition (28) does not depend on $R$ if we assume $P(x,t)$ in the form (31).

The case of anomalous diffusion in a two-layered infinite medium is here derived as a generalization of the normal diffusion. Assuming as usual mass conservation (23), we propose to write the current as follows:

$$J(x,t) = -\frac{\sigma(x)}{2} \frac{\partial}{\partial x} \partial_{0,t}^{1-\alpha(x)} v_\alpha(x) P(x,t), \tag{36}$$

where $\partial_{0,t}^{1-\alpha}$ is the Riemann-Liouville pseudo-differential operator[1] and $v_\alpha(x) = \dfrac{\sigma(x)}{\tau(x)^{\alpha(x)}}$. Expression (36) could be regarded as a generalized fractional Fick's law[2] (Metzler, Klafter, 2000a). As we shall see below, this assumption leads to an analytical formulation which is in good agreement with the Monte Carlo results. Analogously as in the previous case, we assume for sake of simplicity that the parameters characterizing the physical properties are piecewise constant in the two layers, so that $\sigma(x) = \sigma_1, \tau(x) = \tau_1, \alpha(x) = \alpha_1, v_{\alpha,1}(x) = v_{\alpha,1}$ for $x < x_d$ and $\sigma(x) = \sigma_2, \tau(x) = \tau_2, \alpha(x) = \alpha_2, v_{\alpha,2}(x) = v_{\alpha,2}$ for $x > x_d$. Substituting (36) in (23) yields

$$\frac{\partial}{\partial t} P(x,t) = \frac{\partial}{\partial x} \frac{\sigma(x)}{2} \frac{\partial}{\partial x} \partial_{0,t}^{1-\alpha(x)} v_\alpha(x) P(x,t), \tag{37}$$

which, when the medium is homogeneous, takes the form of the well-known Fractional Diffusion Equation (FDE) with generalized diffusion coefficient $D_a = \dfrac{\sigma^2}{2\tau^\alpha}$ (Kilbas et al., 2006; Podlubny, 1999; Oldham, Spanier, 1974; Miller, Ross, 1993; Mainardi et al., 2005; Gorenflo et al., 2004; Gorenflo et al., 2002; Metzler, Klafter, 2000a; Metzler, Klafter, 2004). Moreover, we remark that the inner $\dfrac{\partial}{\partial x}$ in (37) operates on a quantity which has the dimensions of the product of a speed times a concentration, similarly as in the case of Fickian diffusion. Exactly as in the case of normal diffusion, the FDE equation is strictly valid only far in time and in space from the source and therefore it suitably describes the asymptotic behaviour of the walkers with respect to the exact CTRW transport formulation (Scalas et al., 2004). Again, the problem may be decomposed on two separate domains, where

$$\frac{\partial}{\partial t} P(x,t) = \frac{\sigma_1 v_{\alpha,1}}{2} \partial_{0,t}^{1-\alpha_1} \frac{\partial}{\partial x} \frac{\partial}{\partial x} P(x,t) \qquad \text{for } x < x_d \tag{38}$$

$$\frac{\partial}{\partial t} P(x,t) = \frac{\sigma_2 v_{\alpha,2}}{2} \partial_{0,t}^{1-\alpha_2} \frac{\partial}{\partial x} \frac{\partial}{\partial x} P(x,t) \qquad \text{for } x > x_d, \tag{39}$$

with the following conditions at the interface $x_d$:

---

[1] By definition, $\partial_{0,t}^{-\alpha} \{f(x,t)\} \equiv \dfrac{1}{\Gamma(\alpha)} \dfrac{\partial}{\partial t} \displaystyle\int_0^t \dfrac{f(x,t')}{(t-t')^{1-\alpha}} dt'$ for any sufficiently well-behaved function $f$ (See e.g. Kilbas et al., 2006; Podlubny, 1999).

[2] A similar approach may be adopted to describe also superdiffusive transport and in this case it would lead to a non-local fractional Fick's law for the walkers (Paradisi et al., 2001).

$$-\frac{\sigma_1 v_{\alpha,1}}{2}\partial_{0,t}^{1-\alpha_1}\frac{\partial}{\partial x}P(x,t)\bigg|_{x_d^-} = -\frac{\sigma_2 v_{\alpha,2}}{2}\partial_{0,t}^{1-\alpha_2}\frac{\partial}{\partial x}P(x,t)\bigg|_{x_d^+} \qquad (40)$$

and $v_{\alpha,1}\partial_{0,t}^{1-\alpha_1}P(x,t)\big|_{x_d^-} = v_{\alpha,2}\partial_{0,t}^{1-\alpha_2}P(x,t)\big|_{x_d^+}$. (41)

Boundary conditions problems for Fractional Diffusion Equations have been carefully and thoroughly examined by several authors (see e.g. Metzler, Klafter, 2000b; Metzler, Klafter, 2000a; Krepysheva et al. 2006a; Krepysheva et al., 2006b) both for absorbing and reflective boundaries. Yet, to the authors' best knowledge, the analytical approach to subdiffusion through a two-layered medium and its connection with the underlying Monte Carlo microscopic dynamics are new and unexplored subjects. Boundary conditions (40) and (41) generalize expressions (28) and (29). The fundamental solution of the Cauchy problem for (37) when the medium is homogeneous and we require $P(\pm\infty,t)=0$ is given by the Fox's $H$ function, so that $P(x,t)=H(x,t)$ (Metzler, Klafter, 2000a; Metzler, Klafter, 2004). Thus, we propose to generalize solution (31) for the two-layered medium by applying the method of images to the Fox's function. Then, defining $D_{\alpha,i} = \frac{\sigma_i v_{\alpha,i}}{2}$, $i=1,2$, and requiring $D_{\alpha,1} > D_{\alpha,2}$, the Laplace transform of $P(x,t)$ reads

$$P(x,u) = H_1(x,u)\big|_{x<x_d} + R_\alpha(u)H_1(x-2x_d,u)\big|_{x<x_d} + (1-R_\alpha(u))H_2(x+(\beta_\alpha(u)-1)x_d,u)\big|_{x>x_d}. \qquad (42)$$

Here we have denoted the Laplace transform of $P$ by its argument $u$ and the index of the $H$ functions indicates whether the parameters of the right or left layer are to be chosen. In order to satisfy boundary condition (41), the coefficients $\beta_\alpha(u)$ and $R_\alpha(u)$ (which generalize $\beta$ and $R$ appearing in (32) and (33), respectively) are:

$$\beta_\alpha(u) = \sqrt{\frac{D_{\alpha,2}u^{\alpha_1}}{D_{\alpha,1}u^{\alpha_2}}} \quad \text{and} \quad R_\alpha(u) = \frac{\sqrt{(u\tau_1)^{\alpha_1}} - \sqrt{(u\tau_2)^{\alpha_2}}}{\sqrt{(u\tau_1)^{\alpha_1}} + \sqrt{(u\tau_2)^{\alpha_2}}}. \qquad (43)$$

Note that the inverse transform of (42) is given by the convolution of the $H_i$ functions with $R_\alpha(u)$. As in the normal diffusion case, condition (40) does not depend on $R_\alpha(u)$, provided that $P(x,u)$ has the functional form (42). As a particular case, when $\alpha_1 = \alpha_2 = \alpha$, $\beta_\alpha$ and $R_\alpha$ do not depend on $u$ and may be identically rewritten as

$$\beta_\alpha = \sqrt{\frac{D_{\alpha,2}}{D_{\alpha,1}}} \quad \text{and} \quad R_\alpha = \frac{\sqrt{\tau_1^\alpha} - \sqrt{\tau_2^\alpha}}{\sqrt{\tau_1^\alpha} + \sqrt{\tau_2^\alpha}}, \qquad (44)$$

which have a similar structure as the corresponding $\beta$ and $R$ of the normal diffusion case. Moreover, when $\alpha \to 1$ the expressions for normal diffusion are recovered.
As a general remark, the analytical formulation of the transport across a discontinuity, both for anomalous and normal diffusion, is expected to be valid only in the so-called "diffusion limit" approximation, i.e. $t \gg \tau$ and $|x| \gg \sigma$ for each layer separately. This is a consequence of the fact that both the normal diffusion equation and the FDE are asymptotic approximations of the exact CTRW transport model, as mentioned before (Metzler, Klafter, 2000a; Metzler, Klafter, 2004; Scalas et al., 2004; Margolin, Berkowitz, 2004; Marseguerra, Zoia, 2006).

## 6. Results

By taking inspiration from the usual Monte Carlo and analytical approaches to transport phenomena, throughout this Section we assume that the Monte Carlo simulation parameters adopted in the scheme of Section 3 coincide with the coefficients denoted by the same symbols appearing in the analytical formulation of Section 5, for both normal and anomalous diffusion. Then, we proceed to compare the Monte Carlo simulation results for resident concentration and net current with the corresponding analytical curves, under different physical assumptions, leading to different choices of the parameters. As said before, the boundary conditions adopted in (24) and in (37) are those which can best take into account the interface conditions implicitly assumed in the Monte Carlo formulation. In all the following Figures, the analytical curves are plotted as solid lines, whereas Monte Carlo results as dots. The larger statistical fluctuations of the net current $J(x,t)$ with respect to those of the concentration $P(x,t)$ are due to the fact that $J$ is obtained as a difference between forward and backward contributions.

We begin with the normal diffusion case. Concerning the Monte Carlo simulation, we assume at first that walkers maintain their local velocities while crossing the boundary. This assumption is motivated by the fact that in Nature diffusing particles are often found to conserve their speed while crossing an interface separating two media with different physical properties, e.g. in the case of neutrons (Weinberg, Wigner, 1958). By making use of expression (17), this condition imposes the equality of the parametric velocities, namely $\frac{\sigma_1}{\tau_1} = \frac{\sigma_2}{\tau_2}$. Within the analytical approach, this assumption on the microscopic behaviour of the single walkers leads to a continuity of macroscopic resident concentration at the interface, namely $P(x,t)|_{x_d^-} = P(x,t)|_{x_d^+}$ (see eq. (29)). Figures 1a and 1b report resident concentration $P(x,t)$ and current $J(x,t)$ as obtained from Monte Carlo simulation and from analytical formulation as given by (31) and (25). The parameters were set as follows: $x_d$=5, $\tau_1$=0.1, $\tau_2$=0.01, $\sigma_1$=0.707, $\sigma_2$=0.0707, so that $D_1$=2.5 and $D_2$=0.25. It appears that the two approaches are in good agreement.

Any other choice of the parameters implying different parametric velocities (and different local velocities, as well) on the two sides of the interface will give rise to neat bumps in resident concentration profiles: this observation is substantiated by some physical evidences pointing out that macroscopic jumps in resident concentration are possible, even if not so common, e.g. in contaminant transport in test columns (Van Genuchten et al., 1984; Van Genuchten, Parker, 1984; Schwartz et al., 1999; Novakowski, 1992a; Novakowski, 1992b; Leij et al., 1991) or even in temperature profiles (Barrat, Chiaruttini, 2003). An example of resident concentration and current as obtained from Monte Carlo simulation with parameters $x_d$=5 $\tau_1$=0.1, $\tau_2$=0.03, $\sigma_1$=0.707, $\sigma_2$=$\sigma_1$/7 (so that $D_1$=2.5, $D_2$=0.17 and $v_1 \neq v_2$) is shown in Figures 2a and 2b together with the corresponding analytical curves as given by (31) and (25). A good agreement between analytical and Monte Carlo results is evident also in this case.

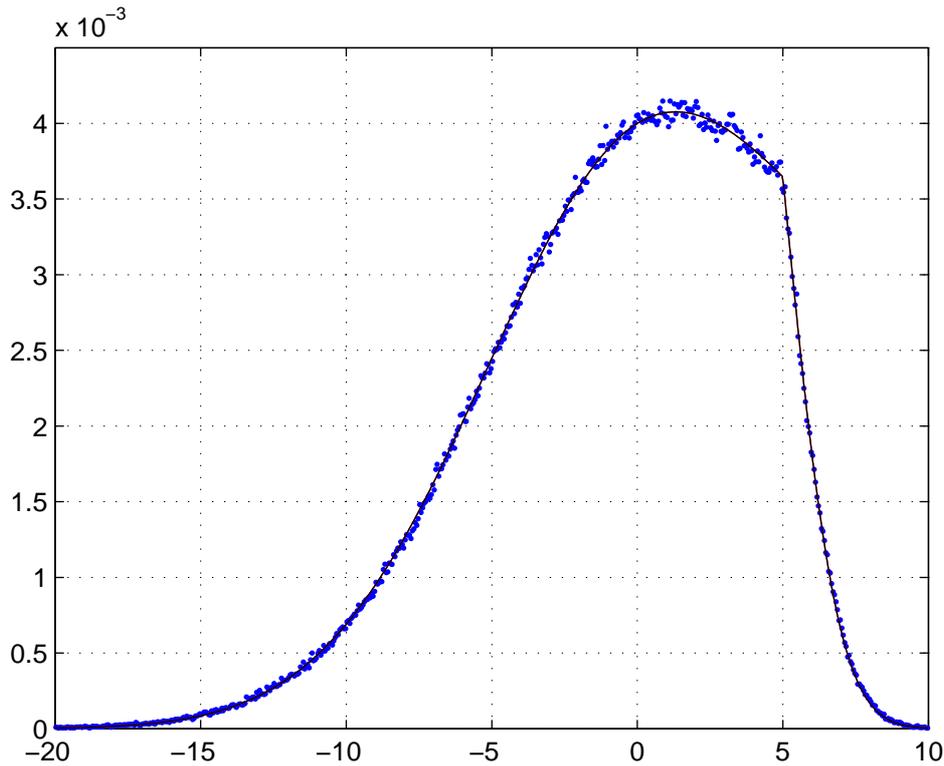

*Figure 1a. Normal diffusion: resident concentration profile P(x,t) at t=6 when local parametric velocities are equal.*

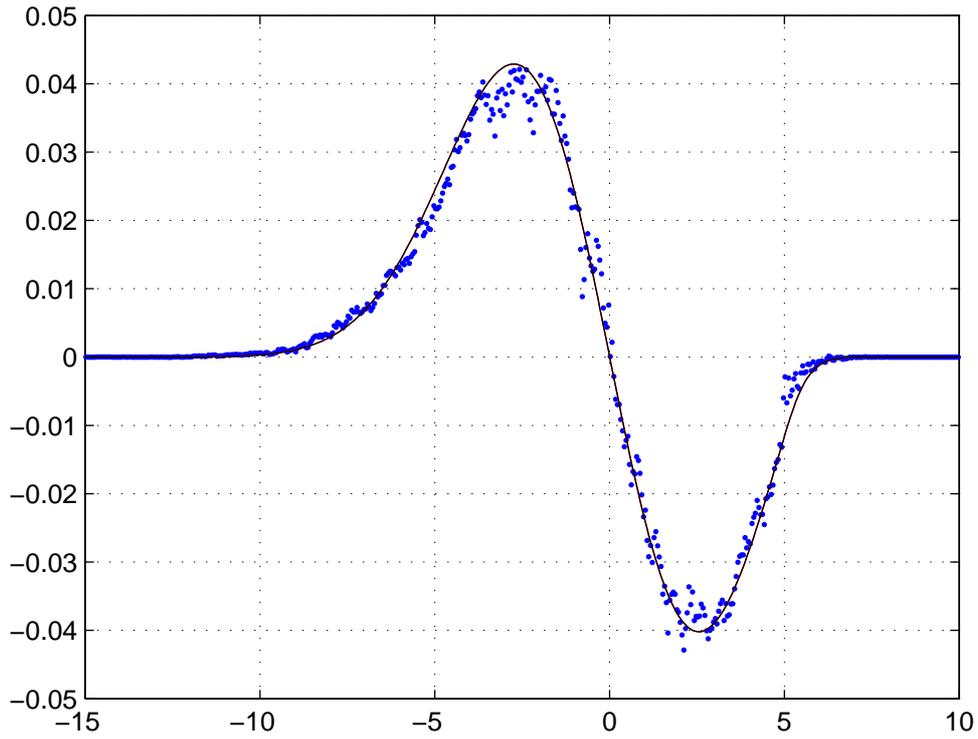

*Figure 1b. Normal diffusion: net current J(x,t) at t=1.5 when parametric velocities are equal.*

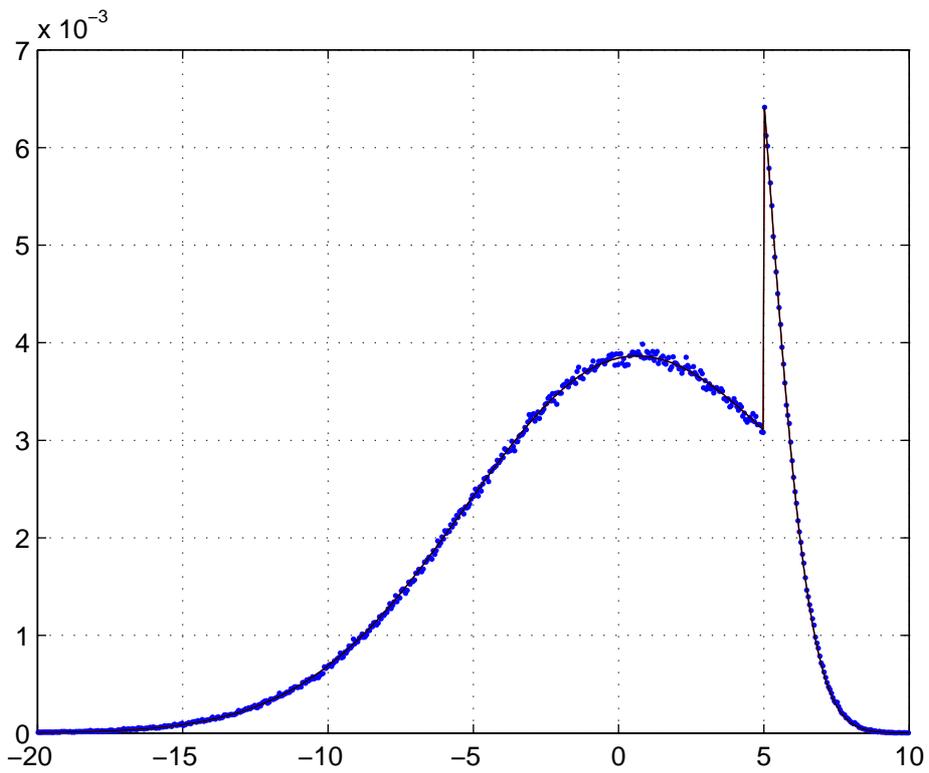

*Figure 2a. Normal diffusion: resident concentration profile P(x,t) at t=6 when parametric velocities are different.*

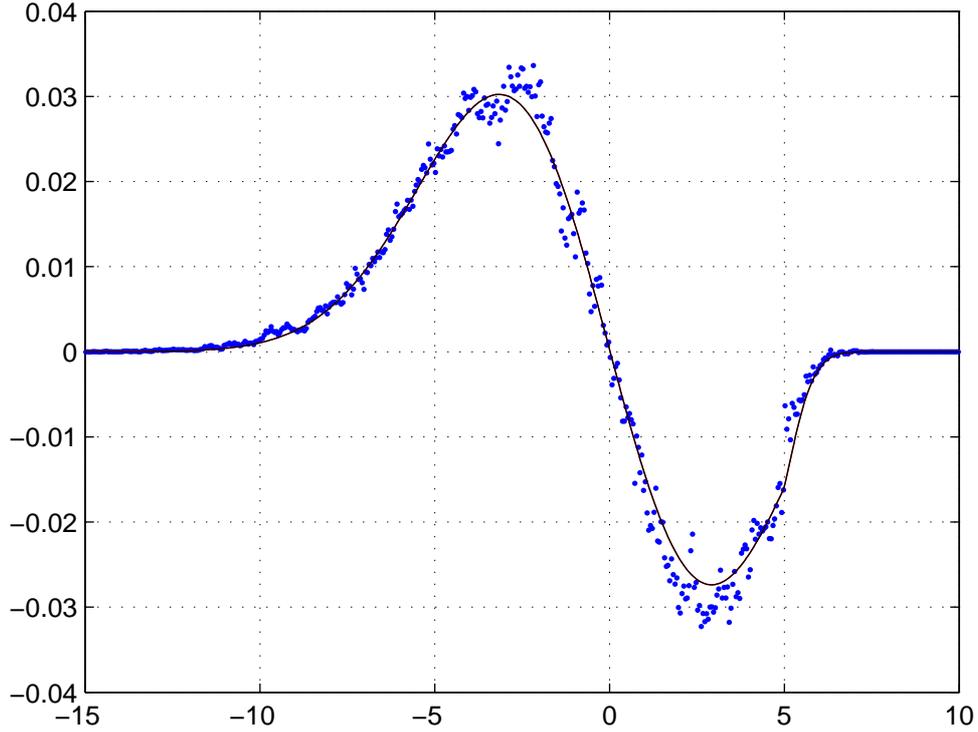

*Figure 2b. Normal diffusion: net current J(x,t) at t=2 when parametric velocities are different.*

As for the anomalous diffusion case, the same physical assumption of particles preserving their local velocities across the boundary imposes two separate constraints on the Monte Carlo parameters, namely the equality of parametric velocities $\frac{\sigma_1}{\tau_1} = \frac{\sigma_2}{\tau_2}$ and also $\alpha_1 = \alpha_2$ (cf. (22)). Differently from the normal diffusion case, these requirements do not automatically imply a smoothly varying resident concentration profile: Figure 3a, where $x_d=5$, $\tau_1=10^{-4}$, $\tau_2=10^{-5}$, $\sigma_1=0.7$, $\sigma_2=\sigma_1/10$, $\alpha_1=\alpha_2=0.5$ (so that $D_{\alpha,1}=24.5$ and $D_{\alpha,2}=0.775$), reveals a neat bump in resident concentration at the interface. Figure 3b shows the net current for the same parameters: the analytical formulations (42) and (36) compare well with the Monte Carlo results, but for slight differences at the interface. With reference to the analytical formulation (42), we remark that imposing resident concentration continuity at the interface requires that the generalized parametric velocities $v_\alpha(x) = \frac{\sigma(x)}{\tau(x)^{\alpha(x)}}$ should be equal on both sides and $\alpha_1 = \alpha_2$. If the same choice is adopted within Monte Carlo simulation, resident concentration profile is shown to smoothly vary across the boundary, while local velocities are not preserved. Figures 4a and 4b compare the Monte Carlo results with the corresponding analytical curves (42) and (36) when the parameters are $x_d=5$, $\tau_1=10^{-3}$, $\tau_2=1.768*10^{-4}$, $\sigma_1=0.4$, $\sigma_2=\sigma_1/4$, $\alpha_1=\alpha_2=0.8$ (so that $v_{\alpha,1}=v_{\alpha,2}=100.5$, $D_{\alpha,1}=20.1$ and $D_{\alpha,2}=5$).

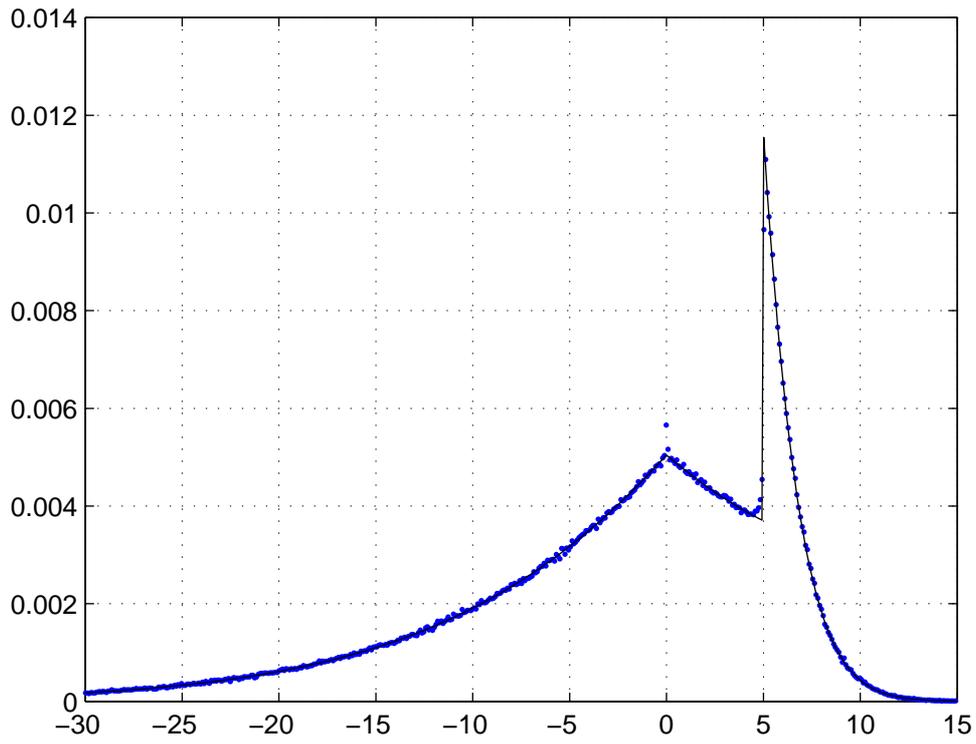

*Figure 3a. Anomalous diffusion: resident concentration profile P(x,t) at t=14 with equal parametric velocities.*

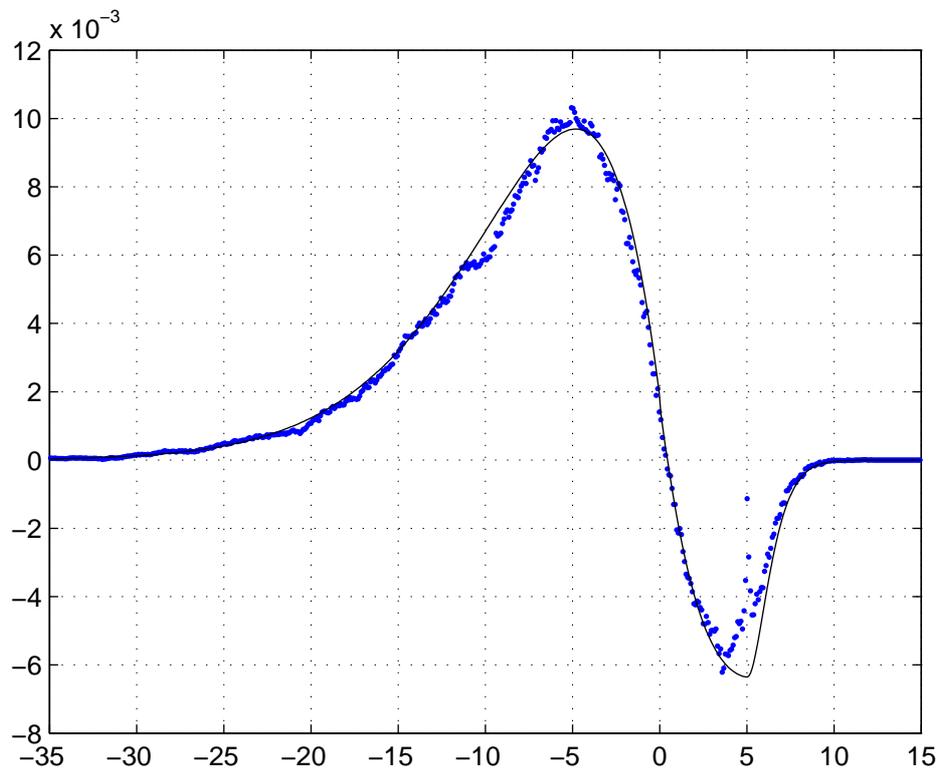

*Figure 3b. Anomalous diffusion: net current J(x,t) at t=1.2 with equal parametric velocities.*

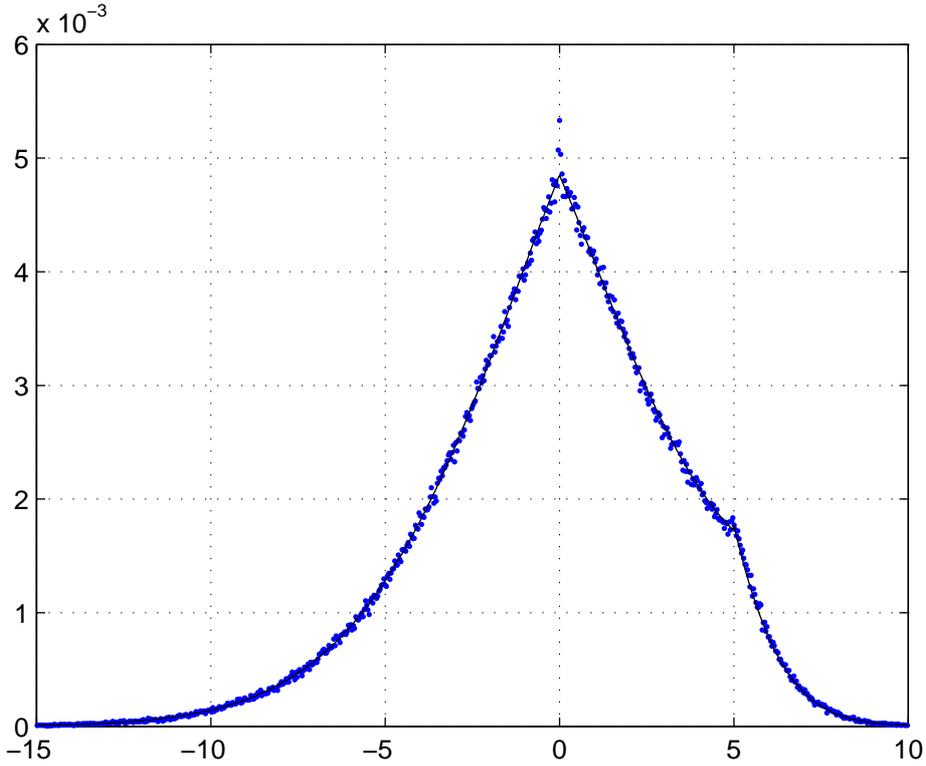
*Figure 4a. Anomalous diffusion: resident concentration profile P(x,t) at t=0.8 with equal generalized velocities.*

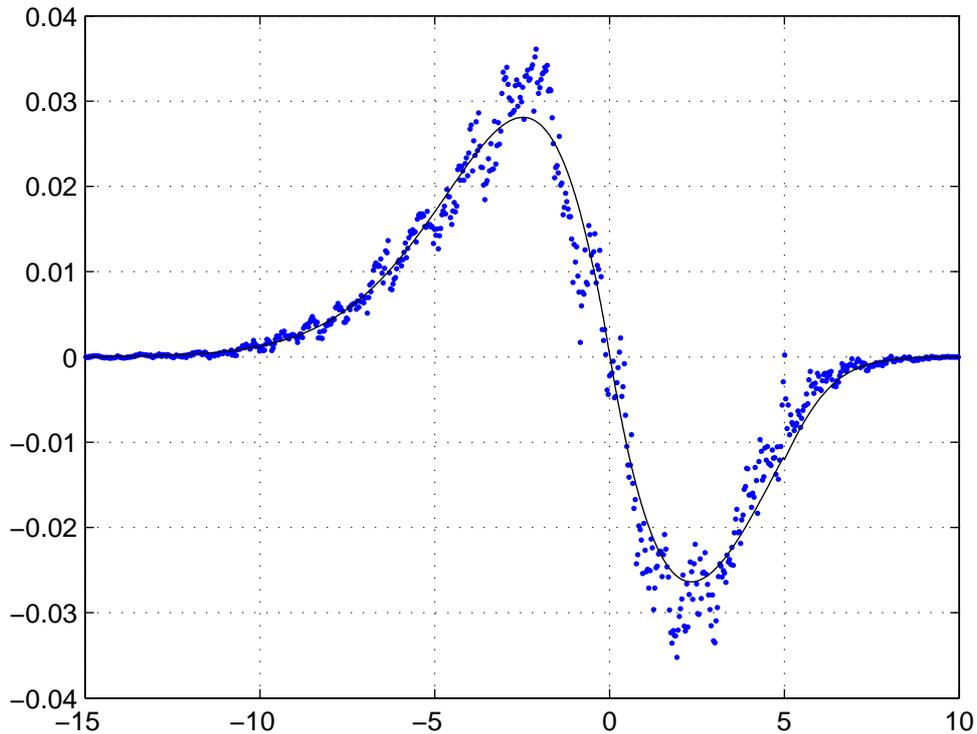
*Figure 4b. Anomalous diffusion: net current J(x,t) at t=0.38 with equal generalized velocities.*

We remark that the previous comparisons between the Monte Carlo simulations and the analytical formulation require that the asymptotic regime, i.e. the diffusion limit, has been attained. This observation is particularly pertinent in correspondence of a bump at the interface: we conjecture that the interface represents a sort of "virtual source" for the walkers and that the attainment of the asymptotic regime must be faster in the Fickian than in the anomalous case. This conclusion is supported by the results shown in Figures 5 and 6, where slight deviations between Monte Carlo and analytical curves at the interface are present only in the anomalous diffusion case (Figure 6) and are completely negligible for normal diffusion (Figure 5). This fact is thought to be related to the "memory effect" which characterizes non-Markovian process such as subdiffusion.

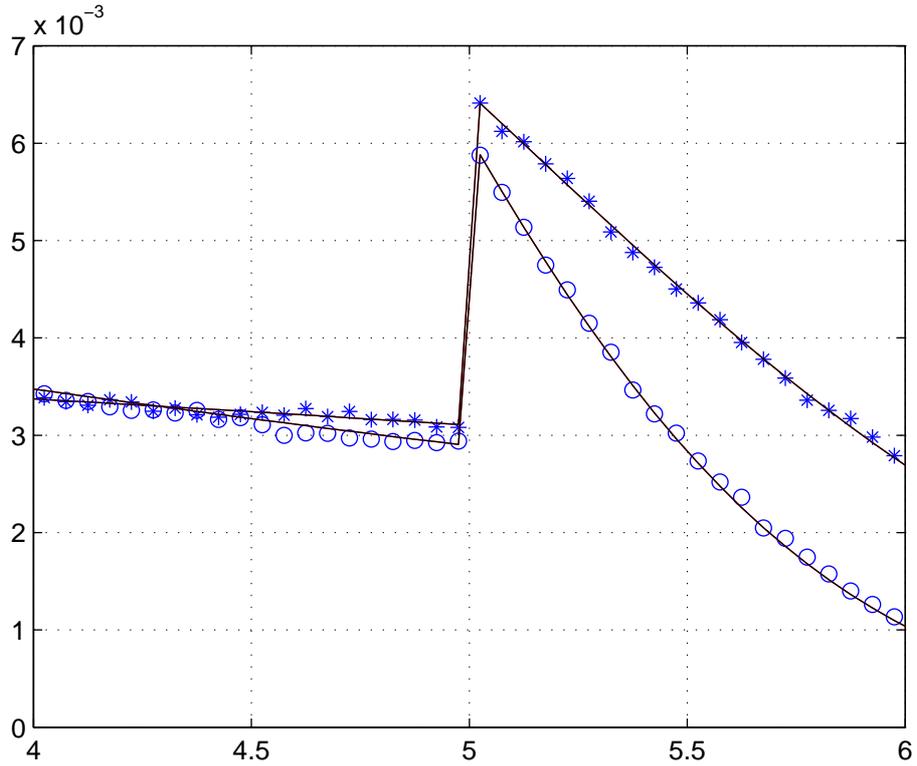

*Figure 5. Normal diffusion: zoom of the bump appearing in Figure 2a at two different times, t=3 (o) and t=6 (*).*

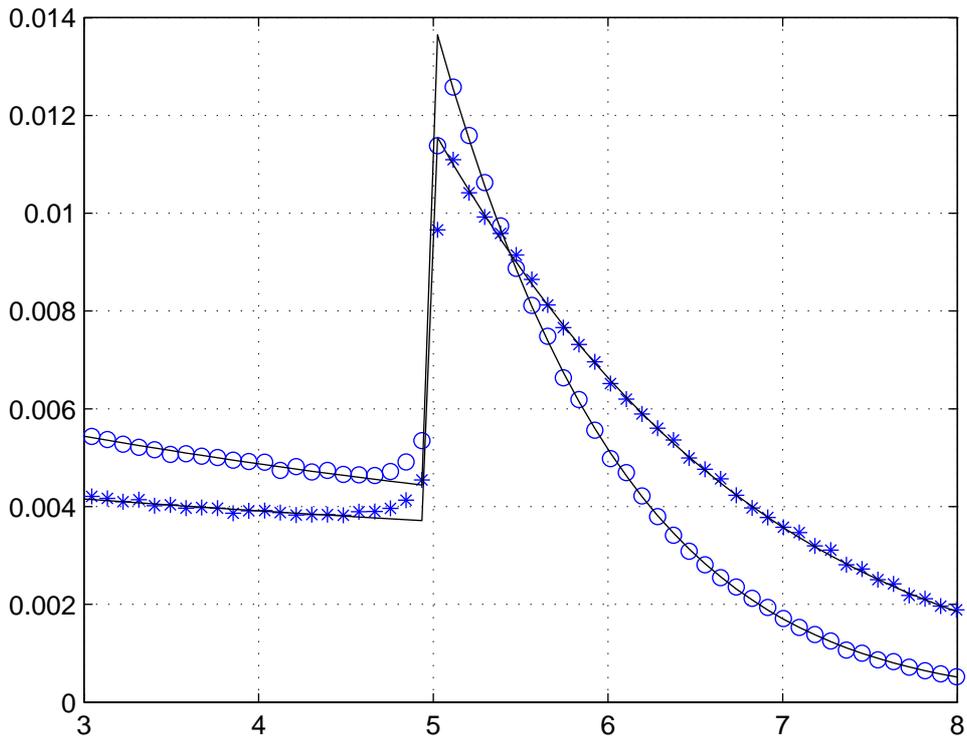

*Figure 6. Anomalous diffusion: zoom of the bump appearing in Figure 3a at two different times, t=2 (o) and t=14(*).*

Finally, note that in the analysis of field or laboratory data the choice of the kind of underlying kinetics (normal or anomalous diffusion regimes) and consequently of the appropriate parameters is usually performed by fitting procedures. However, with reference e.g. to transport of contaminant tracer particles, in most cases resident concentration $P(x,t)$ is not the directly measurable variable, so that experimental evidences are mainly based on breakthrough curves, i.e. on the net current $J(x,t)$ as a function of time collected at the end of an experimental setup (Berkowitz, Scher, 2001; Scher et al. 1991; Berkowitz, Scher, 1998; Berkowitz et al., 2001; Cortis, Berkowitz, 2004; Berkowitz, Scher, 1997; Levy, Berkowitz, 2003; Margolin, Berkowitz, 2002).

## 7. Conclusions

In this paper we have presented a Monte Carlo approach to the transport of particles across an interface separating two media characterized by different physical properties. In particular, this scheme details how to switch from the jump lengths and waiting times distributions of one side to those of the other side. This scheme can take into account both normal and anomalous diffusion by suitably varying the underlying waiting times distribution. Guided by the Monte Carlo results, we have proposed an analytical formulation of the same problem: the most crucial point has been to establish a correspondence between the parameters of the Monte Carlo simulation and the coefficients appearing in the analytical solutions and also between the Monte Carlo rules and the boundary conditions at the interface. For the case of normal diffusion, most physical evidences suggest that particles preserve their local velocities while crossing the interface: we adopted this condition in the Monte Carlo scheme and we derived the corresponding analytical constraints. Under this assumption, resident concentration is found to smoothly vary across the boundary. On the contrary, when the Monte Carlo parameters are such that local velocities vary across the interface and the coefficients of the companion analytical solution are chosen accordingly, resident concentration is found to present a neat jump: this situation, although less common than the previous one, has been reported to occur in some physical contexts.

As for the anomalous diffusion case, we found that imposing that particles preserve their local velocities while crossing the interface does not automatically lead to a smoothly varying resident concentration profile: instead, a neat bump is evident. In order to obtain a smoothly varying resident concentration profile, the analytical formulation suggests to impose the equality of the generalized velocities across the boundary.

In both Fickian and anomalous diffusion cases, the resident concentration and current profiles given by the analytical and Monte Carlo formulations turn out to be in good agreement for any choice of the parameters.

## Acknowledgements

This work was partially supported by the Italian Ministry of Education, University and Research (MIUR), through a project titled *Transport of toxic and/or radioactive contaminants through natural and artificial porous media: models and experiments*.